\documentclass[aps,article]{revtex4}
\usepackage{amsmath}
\usepackage{graphicx, subfigure}
\usepackage{epsfig}
\usepackage{wrapfig}
\usepackage{amsmath}
\usepackage{amssymb}

\newcommand{\betaL}{\beta_{\mathrm{L}}}
\newcommand{\betaLE}{\beta_{\mathrm{L/E}}}
\newcommand{\betaLC}{\beta_{\mathrm{L}}^{~\mathrm{c}}}

\newcommand{\betaLEC}{\beta_{\mathrm{L/E}}^{~\mathrm{c}}}

\newcommand{\gL}{g_{\mathrm{L}}}
\newcommand{\gE}{g_{\mathrm{E}}}
\newcommand{\gLE}{g_{\mathrm{L/E}}}
\newcommand{\gLC}{g_{\mathrm{L}}^{~\mathrm{c}}}

\newcommand{\gLR}{g_{\mathrm{L}}^{~\mathrm{ref}}}
\newcommand{\gTC}{g_{\mathrm{T}}^{~\mathrm{c}}}

\newcommand{\LI}{\Lambda_{\mathrm{L}}^{\mathrm{imp}}}

\newcommand{\LEI}{\Lambda_{\mathrm{L/E}}^{\mathrm{imp}}}
\newcommand{\gIRFP}{g_{\mathrm{4l}}^{\mathrm{IRFP}}}
\newcommand{\gSD}{g_{\mathrm{SD}}^{\mathrm{c}}}

\begin{document}

\title{One, Two, Zero : Scales of strong interactions}

\author{Maria Paola Lombardo} 
\affiliation{INFN-Laboratori Nazionali di Frascati, I-00044, Frascati (RM), Italy}
\author{Kohtaroh Miura}
\affiliation{Kobayashi-Maskawa Institute for the Origin of Particles and the Universe, Nagoya University, Nagoya 464-8602,Japan }
\author{Tiago Nunes da Silva}
\affiliation{Centre for Theoretical Physics, University of Groningen, 9747 AG, Netherlands}
\author
{Elisabetta Pallante}\affiliation{Centre for Theoretical Physics, University of Groningen, 9747 AG, Netherlands}

\begin{abstract}{We discuss our results 
on QCD with a number of fundamental fermions ranging from zero to sixteen.
These theories exhibit a wide array of fascinating phenomena which have
been under close scrutiny, especially in recent years, first and foremost
is the approach to conformality. To keep this review focused, we have chosen 
scale generation, or lack thereof as a guiding theme, however the discussion
will be set in the general framework of the analysis of the phases
and phase transitions of strong interactions at zero and nonzero temperature. 
}
\end{abstract}
\pacs{PACS numbers: 11.15.Ha,11.25.Hf,12.38.Gc,12.60.Nz,12.38.Aw,12.38.Mh}

\maketitle

\section{Adding matter : QCD with an arbitrary flavor number}

Usual QCD dynamics is characterized by spontaneous symmetry breaking
and dynamical mass generation, with the associated scale $\Lambda_{\mathrm{QCD}}$. 
However, when the number of flavors exceeds a critical
number, an infra-red fixed point (IRFP) appears
and prevents the coupling from growing large
enough to break chiral symmetry. 
The theory is then scale invariant - even conformal invariant. 
In the intermediate region, the coupling `walks'
rather than runs
between two typical scales - this is the phenomena of scale separation for
which our results provide a preliminary evidence. 
From a general field theory viewpoint, the analysis of the phase diagram 
of strong interactions as a function of the number of flavor 
adds to our knowledge of the theoretical basis of strong interactions and
their fundamental mechanisms. From a phenomenological viewpoint,
this study deals with a class of models which might play a relevant role
in model building beyond the standard model (BSM) 
\cite{SCGT,Ph_review,Yamawaki:1985zg,Holdom:1984sk,Akiba:1985rr,Appelquist:1986an},
which explain the origin of mass using strong coupling mechanisms
realized in QCD with a large number of flavors. 
All these topics are under active scrutiny 
both theoretically and experimentally 
\cite{Lat1,Lat2,Lat3,Lat4, Itou:2013ofa,
Appelquist:2011dp,Appelquist:Conformal,Appelquist:2009ka,Appelquist:2010xv,
Deuzeman:2008sc,Deuzeman:2009mh,Deuzeman:2012ee,Deuzeman:2012pv,Miura:2011mc,Miura:2012zqa,
Aoki:2014oha,Aoki:2013zsa,Aoki:2013xza,Aoki:2012eq,
Cheng:2014jba,Cheng:2013eu,Cheng:2011ic,
Hasenfratz:2011xn,Hasenfratz:2010fi,Hasenfratz:2009ea,
Fodor:2011tu,Fodor:2009wk,Ishikawa:2013tua,Ishikawa:2013wf,Iwasaki:2003de,
Finland:MWT,Svetitsky:sextet,Kogut:2010cz,Fodor:2009ar,Fodor:2012ty,
Catterall:MWT,Lucini:MWT,
DelDebbio:2010jy,DelDebbio:2010ze,deForcrand,
BraunGies,EKMJ,Alho:2012mh,Gursoy:2010fj,Alho:2013dka,Liao:2012tw,
Ryttov:2012nt,Ryttov:2007cx,Dietrich:2006cm,Antipin:2012sm,
Matsuzaki:2013eva,Matsuzaki:2012xx}

\subsection{Conformality}
Conformal invariance is anticipated to emerge
in the non-Abelian gauge theory with many species (flavors) of fermions
\cite{Caswell:1974gg,Banks:1981nn,Appelquist,Miransky:1997,Appelquist:1999hr}.
This is due to the IRFP
for $N_f > N_f^*$ at a coupling which is not 
strong enough to break chiral symmetry: 
a second zero of the two-loop
beta-function of a non-Abelian gauge theory
implies, at least perturbatively,
the appearance of IRFP conformal symmetry
\cite{Caswell:1974gg,Banks:1981nn}.
In color SU($3$) gauge
theory with $N_f$ massless fundamental fermions,
the second zero appears at $N_f\gtrsim 8.05$,
before the loss of asymptotic freedom (LAF) at
$N_f^{\mathrm{LAF}}=16.5$.
Analytic studies of the conformal transition of strong interactions
have produced a variety of predictions
for the conformal threshold:
the Schwinger-Dyson approach with
rainbow resummations
\cite{Appelquist,Miransky:1997,Appelquist:1999hr}
or the functional renormalization group method
\cite{BraunGies}
suggest the onset of conformal window around $N_f^* \sim 12$.
An all-order perturbative beta-function
\cite{Ryttov:2007cx}
inspired by the Novikov–Shifman–Vainshtein–Zakharov
beta-function of SQCD \cite{Novikov:1983uc} 
leads to a bound $N_f^* > 8.25$.
Instanton studies at large $N_f$ \cite{Velkovsky:1997fe}
claimed a qualitative change of behaviour at $N_f=6$.
The $N_f^{*}$ has also been estimated
for different fermion representations \cite{Dietrich:2006cm}.
Holographic models for QCD in the Veneziano limit 
find $3.7 < (N_f/N_c)^* < 4.2$ \cite{EKMJ}.

\subsection{Pre-conformality} 

The direct inspection of theories at fixed $N_f$ is often inconclusive,
especially close to the expected threshold $N_f^*$. 

An alternative  approach to establish the existence of the
conformal window is to (try to)
observe directly the approach to conformality by monitoring
the evolution of the  results obtained in the broken phase 
as a function of $N_f$.

Moreover, the pre-conformal dynamics at flavor numbers just before
the onset of conformal invariance might serve as a paradigm
for the BSM model buildings that invokes non-perturbative
mechanisms of electroweak symmetry breaking
\cite{Yamawaki:1985zg,Holdom:1984sk,Akiba:1985rr,Appelquist:1986an}.
In such pre-conformal region,
the coupling should vary little -- 
should {\em walk} -- with the scale, 
at variance with the familiar running of QCD.
One important question, of genuinely theoretical nature, 
is to establish the existence and uncover the 
properties of this new class of strongly interacting (quasi) conformal theories.
Because of this, 
the sub--critical region, 
when $N_f$ gets closer and closer to $N_f^*$, 
is interesting per se.~\cite{Ph_review}.
In our study, such pre-conformal dynamics could manifest itself 
either with a clear observation of a separation of scales,
or with a manifestation of a critical behaviour when approaching
$N_f^*$. One possibility is to observe the Miransky-Yamawaki essential 
singularity~\cite{Miransky:1997}.
Alternatively, in an FRG approach~\cite{BraunGies}, 
the pseudo-critical line is almost linear with $N_f$ 
for small $N_f$, and displays a singular behaviour when
approaching $N_f^*$, which could be the only observable
effects, beyond Miransky scaling.
A {\em jumping} scenario in 
which the change from a QCD dynamics to
the conformal window is abrupt is also a distinct possibility
\cite{Antipin:2012sm}.

\subsection{The high temperature path to conformality}

Chiral symmetry is restored at
high temperatures -- in the so-called quark-gluon plasma (QGP)
phase. Both physics intuition and phenomenological analysis
based on functional renormalization group~\cite{BraunGies} 
and finite temperature holographic QCD \cite{Alho:2012mh}
indicate that the conformal phase of cold, many flavor QCD and 
the high temperature chirally symmetric phase are continuously connected. 
In particular, the onset of the conformal window coincides with 
the vanishing of the transition temperature, and the conformal
window appears as a zero temperature limit of
a possibly strongly interacting QGP.

The analysis of the finite temperature phase transition
is a well-established
line of research within the lattice community. In our approach
we build on this experience and use the properties of a thermal
system to learn about general aspects of the phase diagram also
at zero temperature. According to the Pisarski-Wilczek 
scenario~\cite{Pisarski:1983ms}, the most likely possibility for $N_f \ge 3$
is a first order chiral transition in the chiral limit, turning into a 
crossover above a critical mass endpoint, and/or on lattices which are not
large enough. However it should be noted that closer to the conformal
window the dynamics of the light scalar mode might invalidate this simple
picture, and the nature of the thermal transitions poses specific issues. 
We will identify the thermal crossover with confidence for a number of flavors
ranging from four to eight,
and we will complement these results with those
of the deconfinement transition in the quenched model. 
Then, we study the approach to the conformal phase
in the light of the chiral phase transition
at finite temperature with variable number of flavors.
Further, we will argue that
even results in the bare lattice parameters
can be used directly to locate
the critical number of flavors,
thus generalising to finite temperature
the Miransky-Yamawaki phase diagram, Ref. \cite{Miransky:1997}.

\subsection{Setting the scale}

One ubiquitous problem in these studies is the setting of 
a common scale among
theories which are essentially different.
We propose two alternative
possibilities to handle this problem,
one stemming from our own 
work \cite{Miura:2011mc,Miura:2012zqa}, and the other 
from a recent analysis \cite{Liao:2012tw}. 
Interestingly, this latter approach analyses the dependence of 
the confinement parameters on the matter content, and proposes
microscopic mechanisms for confinement
motivated by such $N_f$ dependence.

\subsection{Sketching the phase diagram}
\begin{figure}
\includegraphics[width=8 truecm]{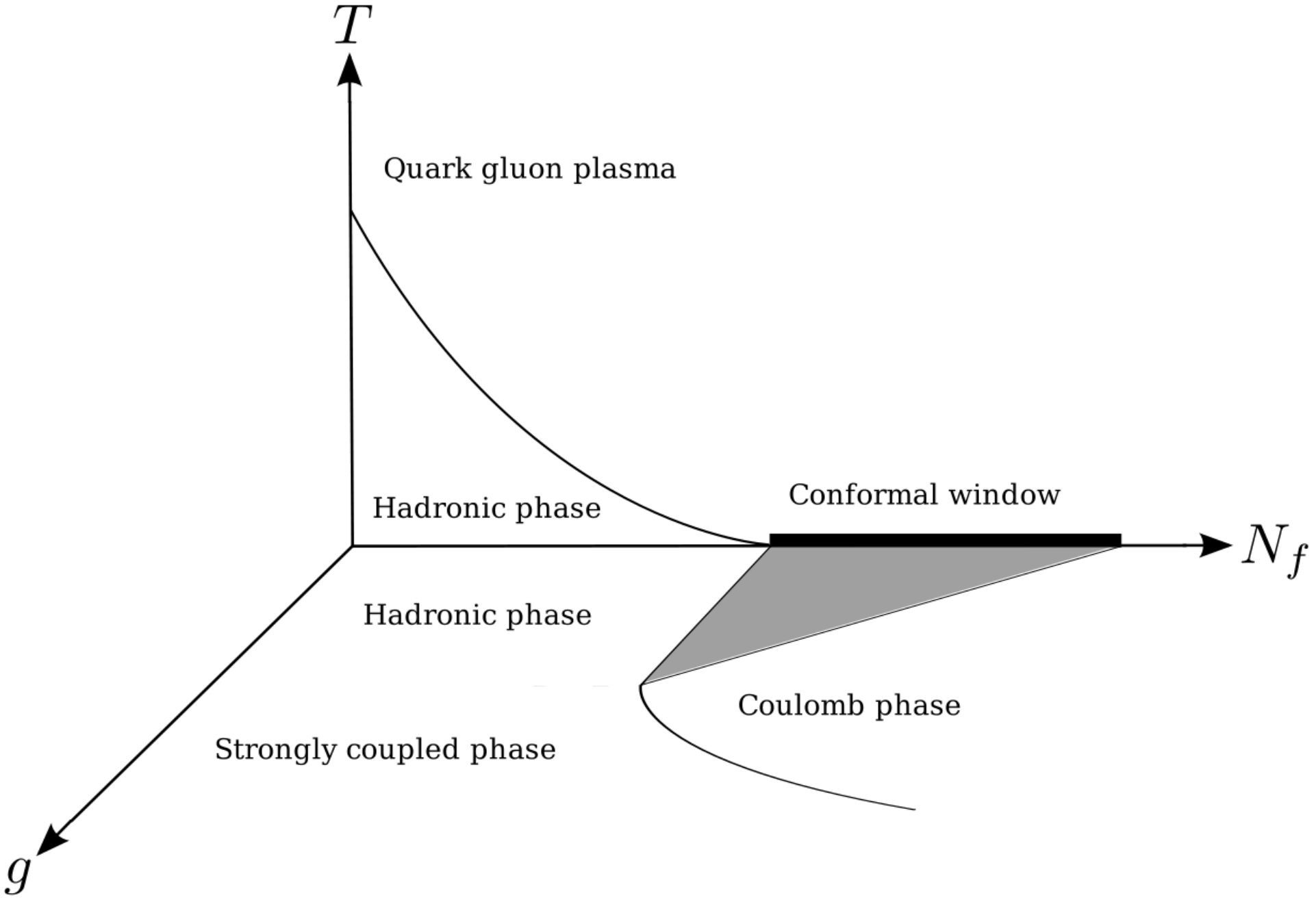}
\caption{\label{Phaseplot} A projected view of the phase diagram of
QCD-like theories in the temperature ($T$), flavor number ($N_f$) and bare
coupling ($g$) space. In the T-$N_f$ plane, the critical line is a phase
boundary between the chirally broken hadronic phase and the chirally
symmetric quark gluon plasma, the zero temperature end point of which is
the onset of the conformal window. The zero temperature projected plane is
inspired by the scenario in Refs.~\protect\cite{Appelquist:1999hr,
Miransky:1997}.}
\label{fig:Combined_phase}
\end{figure}

The phase diagram of QCD emerging from these discussions is sketched in 
Fig.~\ref{fig:Combined_phase}:
the axis is simply the number of light flavors. Ordinary QCD
-- two light flavors -- is marked by an arrow.
The conformal region is on the right hand side, 
and is separated by the essential singularity for a critical number of flavor
(of about eleven according to the current estimates) from the hadronic phase.
The possibility of a first order transition has been discussed as well, and
we will get back to this later in this paper. 
Clearly, as in any system undergoing a phase
transition, the nature and extent of the critical window are purely
dynamical questions whose answer cannot be guessed a priori.
Since the underlying dynamics is completely non-perturbative, lattice
calculations are the only tool to perform an ab initio, rigorous
study of these phenomena, and many lattice studies
have recently appeared~\cite{Lat1, Lat2, Lat3, Lat4}.

We now turn to the presentation of our results. The interested reader
can find all the details in our published papers
\cite{Deuzeman:2008sc,Deuzeman:2009mh,Deuzeman:2012ee,Miura:2011mc,Miura:2012zqa},
and we here omitted many (sometimes important) details for the sake of a more
concise presentation. 
Section \ref{sec:spectrum} is devoted to the strategies
we have used to obtain an (indirect) evidence of conformality
from a direct evidence of chiral symmetry restoration.
Some comments on the mass anomalous dimension associated with our
measurements are included as well.

In Sec.~\ref{sec:Tc},
we investigate the chiral phase transition at finite temperature
for various numbers of flavor, 
and evaluate the onset of the conformal window $N_f^*$
via the vanishing of the transition temperature at large $N_f$.
The results are further exploited in Sec.\ref{sec:QGP}
to highlight a connection between the conformal window and
a {\em cold} QGP.
In Sec.~\ref{sec:scale},
we discuss the issues of scale separation and possible direct evidence,
which are still work in progress.
In Sec.~\ref{sec:summary},
we will summerize our review.

\section{The quest for conformality}\label{sec:spectrum}

In this 
Section, 
we discuss the existence of a conformal phase in $SU(N_c=3)$ gauge
theories in four dimensions.
In this lattice study, we explore the model in the bare parameter space, 
varying the lattice coupling and bare fermion mass.

The analysis of the chiral order parameter and
the mass spectrum of the theory indicates the restoration of chiral symmetry
at zero temperature and the presence of a Coulomb-like phase, 
depicting a scenario compatible with the existence of an
IRFP at nonzero coupling.

Following the T=0 plane of Fig.~\ref{fig:Combined_phase}, 
at a given $N_f>N^*_f$ and
increasing the gauge coupling from $g=0$, one crosses the line of the 
IRFPs, going from a chirally symmetric 
and asymptotically free phase (pre-conformal phase, shaded in the picture) 
to a symmetric, but not asymptotically free one
(Coulomb-like or QED-like phase). A phase transition need not be associated with
the line of IRFPs, differently from what was originally speculated in
Ref.~\cite{Banks:1981nn}. At even larger couplings, a transition to a
strongly coupled chirally asymmetric 
phase will always occur in the lattice regularized theory.
The latter is referred to as a bulk phase transition. 
In the symmetric phases at nonzero coupling the conformal
symmetry is still broken by ordinary perturbative contributions.
They generate the running of the coupling constant which is different on the two
sides of the symmetric phase. See Ref.~\cite{Miransky:1997} for a
detailed discussion of this point. We emphasize that in the region
considered in this paper the conformal symmetry would still be broken by
Coulombic forces.

One lattice strategy to assess conformality was then the following:
first, it was demonstrated that the location of the transition from the chirally
symmetric to the broken phase is not sensitive to the physical temperature
and is therefore compatible with a bulk nature. Subsequently, 
the bare fermion mass dependence of the chiral condensate on
the weak coupling side of the bulk transition clearly favored
a chiral symmetry restoration.
Finally, the behavior of the mass spectrum close to the
bulk transition will be studied, again confirming chiral symmetry restoration
without making use of detailed fits. The mass dependence of the spectrum
allowed the extraction of a candidate mass anomalous dimension.
These results are consistent with the scenario for conformality of 
Fig.~\ref{fig:Combined_phase}.
In the following, we limited ourselves to the presentation of the spectrum
results which are probably those providing a cleanest visual evidence,
and which have been updated and expanded very recently.

All our simulations use 
staggered fermions (Kogut-Susskind)
in the fundamental representation in color $SU(N_c=3)$. 
Here we used a tree level Symanzik improved gauge action to suppress lattice
artifacts, and staggered fermions with the Naik
improvement scheme, that effectively extends the Symanzik improvement to
the matter content.

\subsection{Spectrum}

It has been noted in the past that
one can devise robust signatures of chiral symmetry based
on the analysis of the spectrum results. 
One first significant spectrum observable is the ratio $m_\pi/m_\rho$, 
between the mass of the lightest pseudoscalar state (pion) $m_\pi$
and the mass of the lightest vector state (rho) $m_\rho$. 
In real-life QCD at zero temperature, chiral symmetry is spontaneously broken
and the pion is the (pseudo)Goldstone boson of the broken symmetry, 
implying that its mass will behave as $m_\pi\sim \sqrt{m}$. 
In contrast,
chiral symmetry is restored in the continuum limit in the conformal window.
At the IRFP and at infinite volume, the quark mass dependence of all hadron masses
in the spectrum is governed by conformal symmetry: 
at leading order in the quark mass expansion all masses follow
a power-law with common exponent determined by the anomalous dimension
of the fermion mass operator at the IRFP. 
Hence we expect a constant ratio. Away from the IRFP, 
for sufficiently light quarks and finite lattice volumes, 
the universal power-law dependence receives corrections, 
due to the fact that the theory is interacting but no longer conformal.

\begin{figure}
\begin{minipage}[htb]{\linewidth}
\begin{center}
\includegraphics[width=10cm]{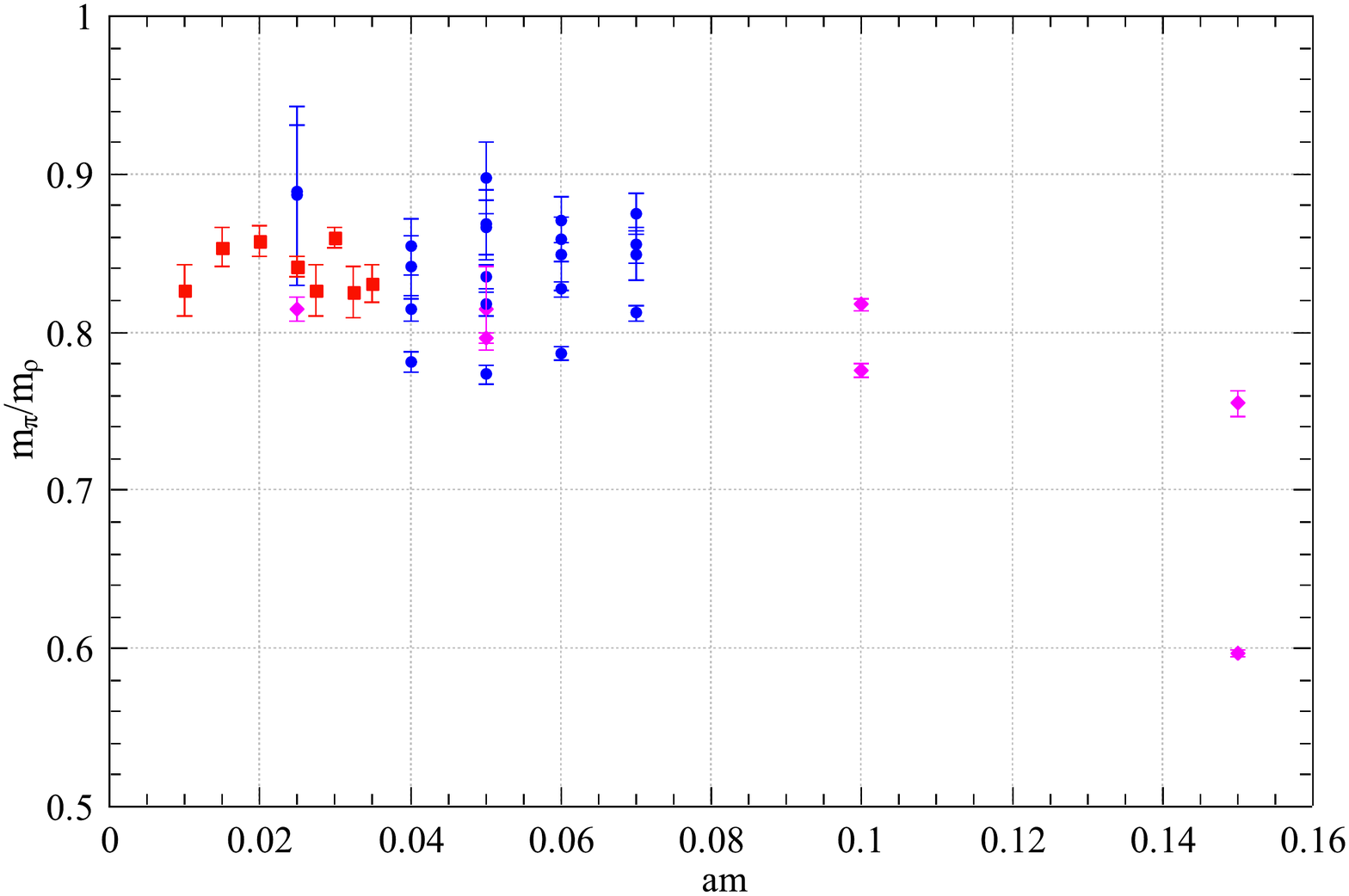}
\caption{Ratio $m_\pi/m_\rho$ as a function of the bare quark mass
for all existing data for $N_f=12$, 
and $N_f=16$: $N_f=12$ data from \protect\cite{Fodor:2011tu} (red squares), 
our $N_f=12$ data and $\beta_L=3.8,3.9,4.0$ (blue circles), 
$N_f=16$ data from \protect\cite{Damgaard:1997ut} (magenta diamonds).}
\label{fig:mpimrhoRatio}
\end{center}
\end{minipage}
\hspace{0.2cm}
\begin{minipage}[htb]{\linewidth}
\begin{center}
\includegraphics[width=10cm]{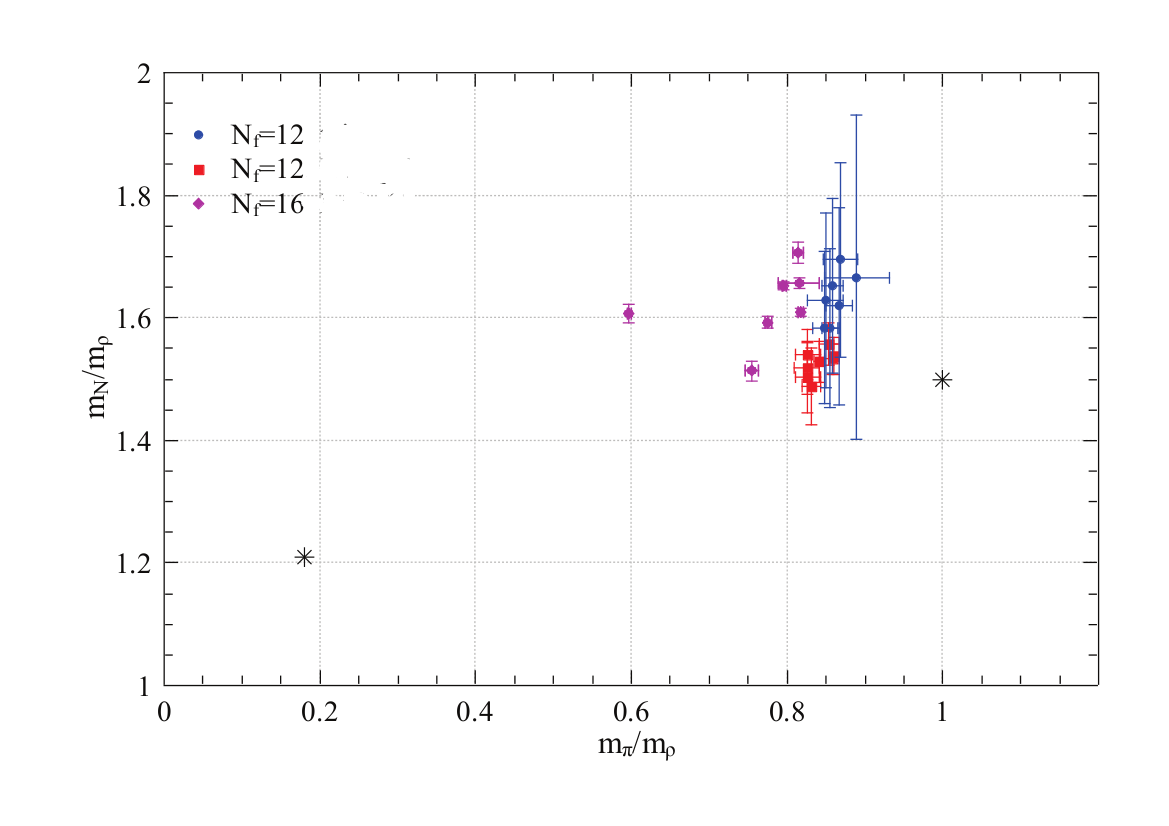}
\caption{Edinburgh plot: $N_f=12$ data from \protect\cite{Fodor:2011tu} (red squares), 
our $N_f=12$ data   and $\beta_L=3.8,3.9$ (blue circles), 
$N_f=16$ data from \protect\cite{Damgaard:1997ut} (magenta diamonds). 
The QCD physical point (black star, leftmost) and 
the heavy quark limit (free theory) point (black star, rightmost) are shown.}
\label{fig:Edplot}
\end{center}
\end{minipage}
\end{figure}

The behaviour of the ratio is demonstrated in Fig. \ref{fig:mpimrhoRatio}:
a conformal scenario seems favoured in the range of masses we are
exploring. Note that the $m_\pi/m_\rho$ 
ratio should go to zero in the chiral limit in the broken phase, and
to a constant value if chiral symmetry is restored.

Analogous conclusions can be drawn from the inspection of
so-called an Edinburgh plot Fig.~\ref{fig:Edplot}. 
The difference with the case of ordinary QCD is indeed striking. 
The modest scattering of the data points 
could be ascribed to the deviation from a perfect power law as discussed
above. It would then be of interest to repeat the same plot
for different couplings : at the IRFP it should indeed reduce to a point.

\subsection{Anomalous Dimension}

\begin{figure}
\includegraphics[width=10truecm]{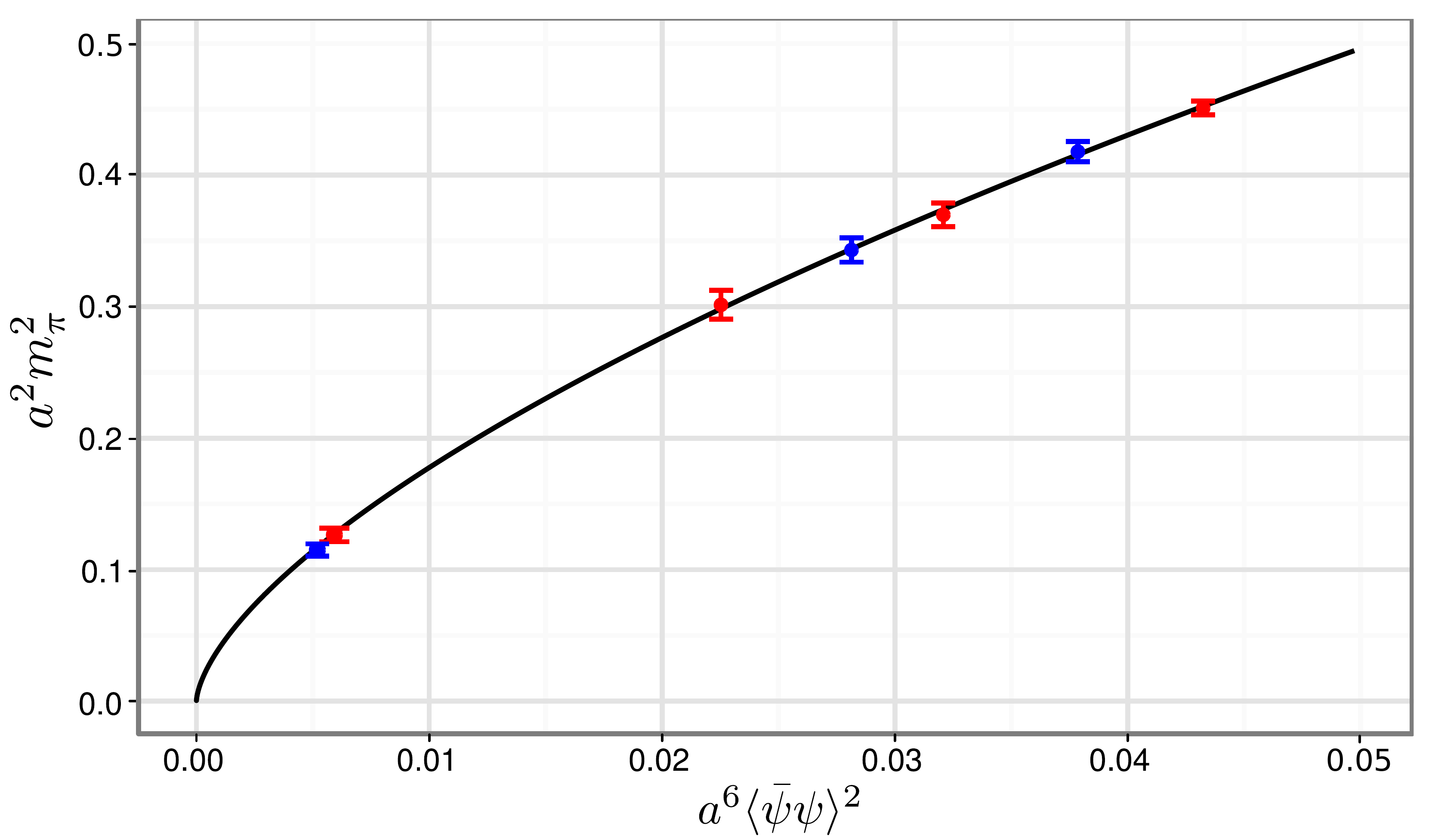}
\caption{\label{fig:pion_pbp_plot} { The relation between the
chiral condensate and the pion mass, for $6/g_L^2$ = 3.9 (blue) and 4.0
(red). The line represents a power law fit to the combined data }}
\end{figure}

To see whether the theory has the anomalous dimension for $N_f = 12$,
we plot the pseudoscalar mass as a function of the
chiral condensate\cite{Kocic:1992is}, as in Fig. \ref{fig:pion_pbp_plot}. 
The data are best fitted by a
simple power-law form 
$(a m_\pi)^2 = A (a^3 \langle\bar\psi\psi\rangle)^{2\delta_\chi}$,
with $\delta_\chi = 0.64(1)$. 
They clearly suggest that chiral symmetry is
restored and that the theory has anomalous dimensions \cite{Kocic:1992is}.
For comparison, in the symmetric
phase and in mean field we expect a linear dependence
with non negative intercept. The presence of anomalous dimensions is
responsible for negative curvature - noticeably opposite to what finite
volume effects would induce - and a zero intercept. The same graph in the
broken phase would show the opposite curvature and extrapolate with a
negative intercept.

\begin{figure}
\includegraphics[width=10truecm]{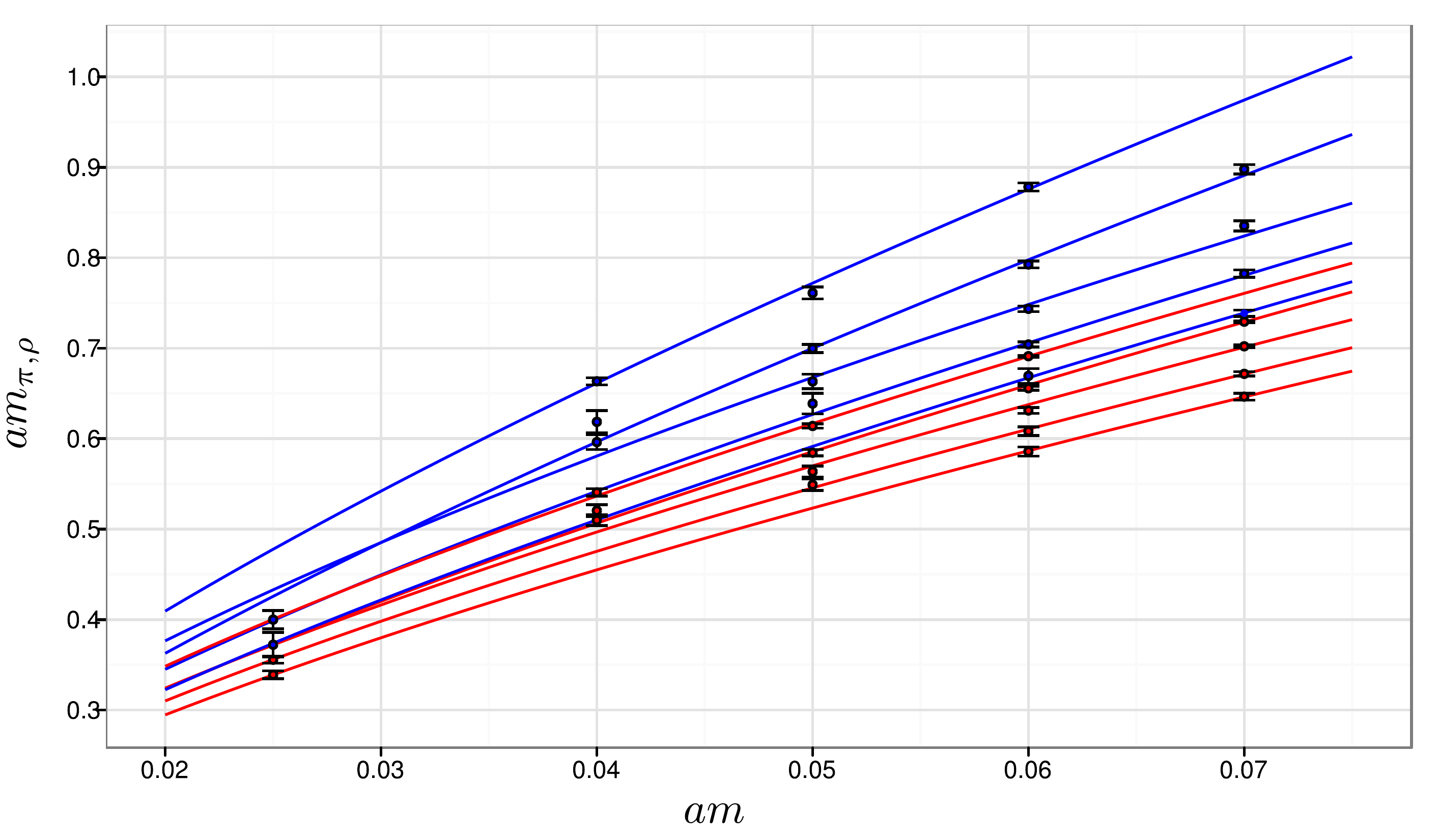}
\caption{\label{fig:meson_mass_plot} {
The relation between
the bare quark mass and the masses of the pion (red) and rho meson (blue),
for $6/g_L^2$ = 3.6, 3.7, 3.8, 3.9 and 4.0 from the uppermost line
down. Power law fits to the separate values of beta are provided.}}
\end{figure}

Further, 
in Fig.~\ref{fig:meson_mass_plot} we report on the measured values of
$m_\pi$ and $m_\rho$ as a function of the bare fermion mass from
our early work \cite{Deuzeman:2009mh}. Here the
lightest point at $am=0.025$ for the vector mass is absent, but a curvature
can still be appreciated. 
Simulations were done on $16^3\times 24$ volumes,
while a set of measurements at larger volumes showed that finite volume
effects were under control. 
The mass dependence
shown in Fig.~\ref{fig:meson_mass_plot} hints again at a few properties of
a chirally symmetric phase. We have fitted both the pion and the
rho mass with a power law ansatz
\begin{equation}
m_{\pi,\rho} = A_{\pi,\rho} m ^{\epsilon_{\pi,\rho}}
\end{equation}
and obtained the results
$A_\pi = 3.41(21)$, $\epsilon_\pi = 0.61(2)$, 
$A_\rho = 4.47(61)$, $\epsilon_\rho = 0.66(5)$ at $6/g_L^2 = 3.9$, and 
$A_\pi = 3.41(21)$, $\epsilon_\pi = 0.61(2)$,
$A_\rho = 4.29(11)$, $\epsilon_\rho = 0.66(1)$ at $6/g_L^2 = 4.0$. 
The accuracies of these fits
are not comparable with those achieved by the fits to the chiral
condensate,
however they allow to draw a few conclusions.
First, the mass dependence of the vector and
pseudoscalar mesons is well fitted by a power-law.
Second, it is also relevant
that the exponents are not unity and $\epsilon_\pi \neq 1/2$. 
The latter result
immediately tells that the pion seen here is not a Goldstone boson of a
broken chiral symmetry. In addition, both mesons have masses scaling with
roughly the same power, as it should be in a symmetric phase, and with
increasing degeneracy towards the chiral limit. The exponent of the power
law being not one, confirms that we are not in the heavy quark
regime. From the results for the exponent $\epsilon$ we can formally extract 
a value for the anomalous dimension consistent with the other lattice results
as well as the analytic estimates \cite{Itou:2013ofa}.
Needless to say, a full control on the systematic and on the corrections to
scaling  is needed  before making such identification with confidence.

\subsection{Inside the conformal phase : lattice matters at strong coupling!}
We have previously discussed 
a strong coupling zero temperature transition -- a bulk transition -- 
within the conformal window. 
However,
if we were to use a perfect action, the conformal phase discussed
above would extend all the way till the infinite coupling limit.
The role of improvement in this case
is really dramatic! A perfect action would destroy a phase transition.
No surprise, of course:
these are strong coupling phenomena taking place away
from the continuum limit, hence extra terms in the actions which are
irrelevant in the continuum might well become relevant.

But then, how would an ordinary improved action (as opposed
to a perfect action) affect the phase transition? The
evidence we have so far is in this case \cite{Deuzeman:2012ee} 
the bulk transition moves
towards stronger coupling (consistently with the fact that
it will eventually disappear with a perfect action),
and a second transition develops. Among these two transitions
we have a phase with an unusual realization of chiral symmetry,
observed also in other studies \cite{Cheng:2011ic}.

From the perspective of the analysis of continuum many flavor QCD,
these observations are just due to a peculiar form of lattice artifacts.
Bulk transitions are however interesting for several 
reasons including fundamental questions in the quantum field theory,
for example the existence of an non--trivial UV 
fixed point in four dimension away from the perturbative domain 
as well as modeling of condensed matter systems, such as graphene,
and the new phases discussed here might well be of interest in
these contexts.

\section{Near-conformal : continuum and lattice}\label{sec:Tc}

In this Section we discuss results for $N_f=0,~4,~6,~8$,
approaching the conformal window from below.
In this case the results have been obtained with
a fixed bare quark mass, and no attempt has been done to
extrapolate to the chiral limit.

In order to monitor the behaviour of these theories we had
to choose an observable, and we set for the (pseudo)critical
temperature. For each $N_f$
results are given for several values of $N_t$: this 
is necessary
in order to control the approach to the continuum limit,
as we will show below.

\begin{table*}
\caption{
Summary of the (pseudo) critical lattice couplings $\betaLC$
for the theories with $N_f=0,~4,~6,~8$, $am=0.02$
and varying $N_t=4,~6,~8,~12$
\protect\cite{Deuzeman:2008sc,Miura:2012zqa}.
}\label{Tab:bc}
\begin{center}
\begin{tabular}{c|cccc}
\hline\hline
$N_f\backslash N_t$ &	
$4$&
$6$&
$8$&
$12$\\
\hline
$0$ &
$7.35\pm 0.05$&		
$7.97\pm 0.07$&	
$8.26\pm 0.06$&		
$-$\\
$4$ &		
$5.65\pm 0.05$&		
$6.00\pm 0.05$&	
$6.15\pm 0.15$&		
$-$\\
$6$ &
$4.675\pm 0.05$&	
$5.025\pm 0.05$&	
$5.20\pm 0.05$&	
$5.55\pm 0.1$\\	
$8$ &
$-$&
$4.1125\pm 0.0125$&
$4.275\pm 0.05$&
$4.34\pm 0.04$\\
\hline\hline
\end{tabular}
\end{center}
\end{table*}

We have used a common bare fermion mass $ma = 0.02$
for all simulations at finite $N_f$.
Introducing a bare fermion mass,
any first order phase transition
will eventually turn into a crossover
for masses larger than some critical mass,
and any second order transition will immediately become
a crossover. 
Since the chiral condensate looks smooth in our results,
we use the terminology of ``chiral crossover'' in the following.
In Table \ref{Tab:bc} we summarize
the (pseudo)critical lattice couplings $\betaLC$ as a function
of $N_f$ and $N_t$
associated with the thermal crossover .
These are our raw data.

\subsection{IRFP from the lattice results}

\begin{figure}
\includegraphics[width=10cm]{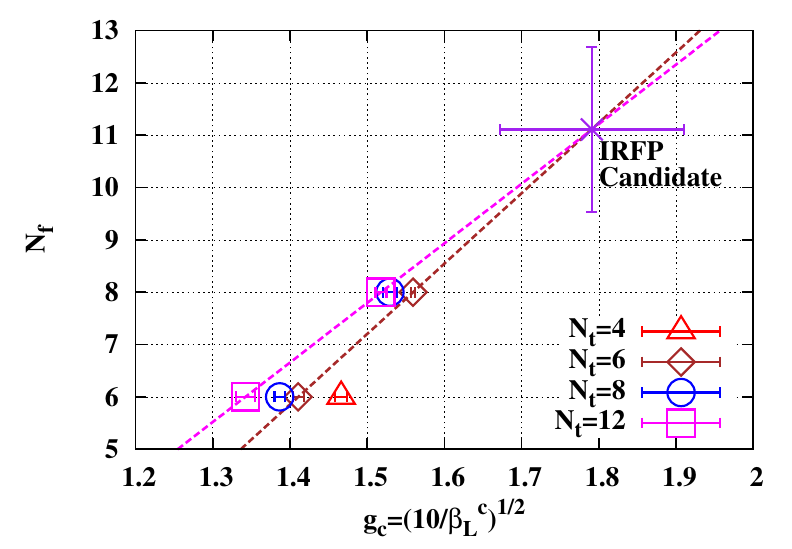}
\caption{(Pseudo) critical values of the lattice coupling
$\gLC=\sqrt{10/\betaLC}$ for theories with $N_f=0,~4,~6,~8$ 
and for several values of $N_t$
in the Miransky-Yamawaki phase diagram.
We have picked up $\gLC$ at $N_f = 6$ and $8$,
and considered ``constant $N_t$'' lines
with $N_t = 6,\ 12$.
{If the system is still described by
one parameter beta-function in this range of coupling,
the IRFP could be located at the intersection of the
fixed $N_t$ lines -- or equivalently, in the region where
the step-scaling function vanishes. To demonstrate the procedure
--as a preliminary example -- 
we have considered the intersection
of the $N_t = 12$ and $N_f = 6$ lines}.}
\label{Fig:MY}
\end{figure}

Let us plot the lattice critical couplings
$\gLC(N_f,N_t) = \sqrt{10/\betaLC(N_f,N_t)}$
(Table \ref{Tab:bc})
in the space spanned by the bare coupling $\gL$
and the number of flavor $N_f$, and 
 consider the lines which connect
$\gLC$ with $N_t$ fixed: $\gLC(N_f)|_{N_t=\mathrm{fix}}$
(see Fig.~\ref{Fig:MY}).
These pseudo-critical thermal lines 
separate a phase where chiral symmetry
(approximately) holds
from a phase where chiral symmetry is spontaneously broken
\footnote{
It would be of interest to study the interrelation of such lines
with the zero temperature first order transition line
observed in the conformal window
\cite{Deuzeman:2012ee,Deuzeman:2009mh,
Cheng:2011ic,Hasenfratz:2010fi,Hasenfratz:2011xn,Damgaard:1997ut,deForcrand}.}.
The resultant phase diagram may be seen
as an extension of the well-known
Miransky-Yamawaki phase diagram \cite{Miransky:1997}
to finite temperature.

We here argue that
the critical number of flavor $N_f^*$ can be read off
from the crossing point of thermal lines obtained for different $N_t$.
To see this,
we consider the well-known step-scaling function: 

\begin{equation}
\Delta\betaL^s = \betaL - {\betaL}^{\prime}
\end{equation}

where $\betaL$ and $\betaL^{\prime}$ 
give the same physical scale $\xi$:

\begin{align}
\xi = a(\betaL)\hat{\xi} = a({\betaL}^{\prime})\hat{\xi}^{\prime}
\ .\label{eq:unique_xi}
\end{align}

Here, $\hat{\xi}$ is the dimension-less
lattice correlation length, and 
$\hat{\xi}/ \hat {\xi^{\prime}} = s$.
In our case, $\xi = T_c^{-1}$, $\hat{\xi} = N_t, 
\hat{\xi}^{\prime} = N_t^{\prime}$,
and the above relation Eq.~(\ref{eq:unique_xi}) reads 

\begin{align}
T_c^{-1} = N_t\ a(\betaLC) = N_t^{\prime}\ a({\betaLC}^{\prime})
\ .\label{eq:unique_Tc}
\end{align}

As discussed in the previous study Ref.~\cite{Hasenfratz:2011xn},
$\Delta \betaL^s = 0$ holds at the IRFP
regardless the scale factor $s$.
In principle,
we could then compute the step-scaling function from
our numerical results,
and try to see where it vanishes.
Alternatively,
we can look for the intersection
of pseudo-critical thermal lines:
obviously, $\Delta \betaL^s = 0$ holds at the intersection point 
regardless the value of the scale factor $s$.

To demonstrate this procedure,
we consider
the pseudo-critical lines obtained for $N_t = 6$ and $N_t=12$
as shown in Fig.~\ref{Fig:MY}.
Note their positive slope:
the lattice critical coupling $\gLC$ is
an increasing function of $N_f$.
This is a consequence
of enhanced fermionic screening for a large number
of flavor, as noted first in Ref.~\cite{Kogut:1985pp}.
Interestingly, the slope decreases with increasing $N_t$,
which allows for a crossing point at a larger $N_f$.
Thus, we estimate the intersection at
$(\gLC, N_f^*) = (1.79\pm 0.12, 11.1\pm 1.6)$.

\subsection{Towards the continuum limit: estimating again the 
conformal threshold}

Let us now fix $N_f$ and consider 
the pseudo-critical temperatures $T_c$ in physical units:
\begin{align}
&T_c\equiv \frac{1}{a(\betaLC)\cdot N_t}\ .\label{eq:Tc}
\end{align}

We introduce the normalised critical temperature
$T_c/\Lambda_{\mathrm{L/E}}$ 
(see e.g. \cite{Gupta:2000hr})
where $\Lambda_{\mathrm{L}}$ ($\Lambda_{\mathrm{E}}$)
represents the lattice (E-scheme) Lambda-parameter
defined in the two-loop perturbation theory
with or without a renormalisation group inspired
improvement~\cite{CAllton}.

We consider the two-loop beta-function
\begin{align}
&\beta(g)
=-(b_0 {g}^3 + b_1 {g}^5)\ ,\label{eq:beta_func}\\
&b_0
=
\frac{1}{(4\pi)^2}
\Biggl(
\frac{11C_2[G]}{3}-\frac{4T[F]N_f}{3}
\Biggr)\ ,\label{eq:b0}\\
&b_1
=
\frac{1}{(4\pi)^4}
\Biggl(
\frac{34(C_2[G])^2}{3}
-\biggl(\frac{20C_2[G]}{3}+4C_2[F]\biggr)T[F]N_f
\Biggr)\ ,\label{eq:b1}
\end{align}
with $(C_2[G],\,C_2[F],\,T[F])=(N_c,\,(N_c^2-1)/(2N_c),\, 1/2)$.
The coupling $g$ can be either 
the lattice bare coupling
$\gL = \sqrt{10/\betaL}$ or the E-scheme renormalised coupling
$\gE = \sqrt{3(1-\langle P \rangle(\gL))}$,
where $\langle P\rangle(\gL)$ is the zero temperature
plaquette value. 
If the one-loop perturbation theory exactly holds,
the E-scheme coincides the lattice scheme.

Integrating Eq.~(\ref{eq:beta_func}),
we obtain the well-known two-loop asymptotic scaling relation,
\begin{align}
R(\gLE)\equiv
a(\gLE)\Lambda_{\mathrm{L/E}}
= \bigl(b_0\gLE^2\bigr)^{-b_1/(2b_0^2)}
\exp\biggl[
\frac{-1}{2b_0\gLE}
\biggr]
\ ,\label{eq:RL}
\end{align}
where $\Lambda_{\mathrm{L}}$ ($\Lambda_{\mathrm{E}}$)
is the Lattice (E-scheme) Lambda-parameter. 

To take into account higher order corrections,
we have also considered the renormalisation group inspired
improvement~\cite{CAllton}
\begin{align}
R^{\mathrm{imp}}(\betaLE)=
\LEI~a(\betaLE)
\equiv
\frac{R(\betaLE)}{1+h}
\times
\Biggl[
1 + h\
\frac{R^2(\betaLE)}{R^2(\beta_0)}
\Biggr]\ ,
\label{eq:RL_imp}
\end{align}
where $\betaLE = 10/(\gLE)^2$.
The coupling $\beta_0$ can be arbitrarily set 
and the parameter $h$ is adjusted
so as to minimise the scaling violation.
Note that $h = 0$ reproduces the standard asymptotic scaling law
Eq.~(\ref{eq:RL}).

We now substitute $\betaLEC$
into the temperature definition Eq.~(\ref{eq:Tc}),
and insert the scale $\Lambda_{\mathrm{L/E}}$:
\begin{align}
\frac{1}{N_t}=\frac{T_c}{\Lambda_{\mathrm{L/E}}}
\times \Bigl(\Lambda_{\mathrm{L/E}}~a(\betaLEC)\Bigr)
\ .\label{eq:T_Lam}
\end{align}
Eq.~(\ref{eq:T_Lam}) allows us
to define the (normalised) critical temperature
$T_c/\Lambda_{\mathrm{L/E}}$.
When we adopt the improvement Eq.~(\ref{eq:RL_imp}),
$T_c/\Lambda_{\mathrm{L/E}}$ is upgraded into
$T_c/\LEI$.

\begin{align}
\frac{T_c}{\Lambda_{\mathrm{L/E}}}
=\frac{R(\gLE)}{N_t}
= \bigl(b_0\gLE^2\bigr)^{-b_1/(2b_0^2)}
\exp\biggl[
\frac{-1}{2b_0}
\biggr]
\ ,\label{eq:TcL}
\end{align}

where $\gLE$ denotes either the bare lattice coupling
or the coupling defined in the E scheme. 
In addition, we consider the renormalisation group inspired 
definition,

\begin{align}
\frac{T_c}{\LEI}
= \frac{R^{\mathrm{imp}}(\gLE)}{N_t}
\ ,\label{eq:TcL_imp}
\end{align}

where $R^{\mathrm{imp}}$ is given by Eq.~(\ref{eq:RL_imp}).
The numerical results for
$T_c/\Lambda_{\mathrm{L/E}}$ and $T_c/\LEI$ 
are collected in 
Table \ref{tab:TcL} and Table \ref{tab:TcLE}.

\begin{table*}
\caption{
Summary of
$T_c/\Lambda_\mathrm{L}$ and
$T_c/\LI$ for various $(N_f,N_t)$.
The first (second) line at fixed $(N_f,N_t)$
shows the value of $T_c/\Lambda_\mathrm{L}$ ($T_c/\LI$),
and the last two columns provide
the parameter $h$ and $\beta_0$ appeared
in the improved asymptotic scaling
Eq.~(\protect\ref{eq:RL_imp}).
}\label{tab:TcL}
\begin{center}
\begin{tabular}{c|cccc|cc}
\hline\hline
$N_f\backslash N_t$ &
$4$&
$6$&
$8$&
$12$&
$h$&
$\beta_0$\\
\hline
$0$ &
$18.11\pm 0.65$&
$18.21\pm 0.91$&
$16.56\pm 0.71$&
$-$&
$-$&
$-$\\
\quad &
$16.29\pm 0.75$&
$17.81\pm 1.02$&
$16.56\pm 0.78$&
$-$&
$0.05$&
$8.26$\\
\hline
$4$&
$21.99\pm 1.04$&
$19.98\pm 0.95$&
$17.12\pm 2.43$& 
$-$&
$-$&
$-$\\
\quad &
$16.56\pm 1.44$&
$18.67\pm 1.38$&
$17.12\pm 3.41$&
$-$&
$0.30$&
$6.15$\\
\hline
$6$ &
$25.41\pm 1.43$&
$25.33\pm 1.43$&
$22.94\pm 1.29$&
$22.30\pm 2.52$&
$-$&
$-$\\
\quad &
$21.66\pm 1.64$&
$23.87\pm 1.58$&
$22.21\pm 1.40$&
$22.30\pm 2.66$&
$0.03$&
$5.55$\\
\hline
$8$ &
$-$&
$50.05\pm 0.87$&
$47.06\pm 3.28$&
$34.34\pm 1.91$&
$-$&
$-$\\
\quad &
$-$&
$34.32\pm 1.40$&
$42.67\pm 6.33$&
$34.34\pm 3.90$&
$1.08$&
$4.34$\\
\hline\hline
\end{tabular}
\end{center}
\end{table*}

\begin{table*}
\caption{
Summary of $T_c/\Lambda_\mathrm{E}$
and $T_c/\LEI$ for $N_f = 6$ and $N_f = 8$.
The first (second) line at fixed $(N_f,N_t)$
shows the value of $T_c/\Lambda_\mathrm{E}$ ($T_c/\LEI$),
and the last two columns give
the parameter $h$ and $\beta_0$ appeared
in the improved asymptotic scaling Eq.~(\protect\ref{eq:RL_imp}).
For $N_f = 6$, the improvement was not necessary. 
}\label{tab:TcLE}
\begin{center}
\begin{tabular}{c|cccc|cc}
\hline\hline
$N_f\backslash N_t$ &
$4$&
$6$&
$8$&
$12$&
$h$&
$\beta_0$\\
\hline
$6$ &
$74.22\pm 5.86$&
$75.47\pm 8.17$&
$74.56\pm 9.08$&
$75.13\pm 10.76$&
$-$&
$-$\\
\hline
$8$ &
$-$&
$422.54\pm 23.06$&
$422.61\pm 38.59$&
$316.03\pm 20.06$&
$-$&
$-$\\
\quad &
$-$&
$312.16\pm 33.13$&
$393.58\pm 60.01$&
$316.03\pm 31.52$&
$0.40$&
$4.34$\\
\hline\hline
\end{tabular}
\end{center}
\end{table*}

Let us know consider the results at fixed $N_f$: 
for each $N_f$ , the
ratio in either Table approaches a constant by increasing $N_t$,
enabling us (with the due caveats) to interpret these asymptotic
values as continuum estimates.

Let us then take the values corresponding to the largest $N_t$
and consider their $N_f$ dependence : $T_c/\Lambda$ apparently
increases with $N_f$! How this can be reconciled with a vanishing
$T_c$ in the chiral limit? This is discussed below, and again in the
last Section.

\subsection{The critical number of flavor
and the vanishing critical temperature}
\label{subsec:TcM}

The apparent puzzle above immediately suggests that $\Lambda$
vanishes faster than $T_c$ when approaching $N_f$, i.e.
has a strong sensitivity to the IR dynamics affected by the conformal
threshold. 

To observe the vanishing of $T_c$ we then need to properly define
a UV reference scale. Here we will review our first attempt
to do so which relies heavily on perturbation theory, while
in the last Section we will describe our ongoing work
on this subject. 

\begin{figure}
\begin{center}
\includegraphics[width=10cm]{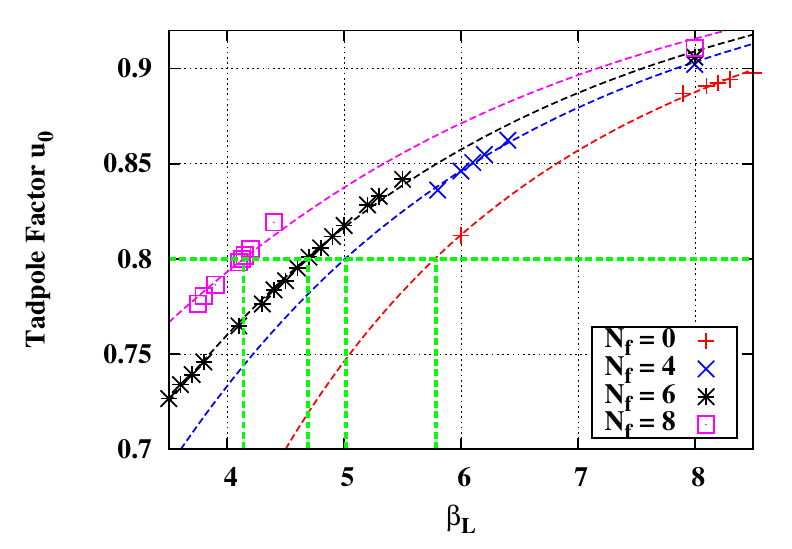}
\caption{
The $\betaL$ dependence of the tadpole factor $u_0$
at zero temperature ($12^4$ lattice volume).
{At each $N_f$,
the dashed line represents the fit for data
with the ansatz $u_0 = 1 - A/(1 + B\cdot\betaL^2)$.}
We consider a constant $u_0$ ({\em e.g.} $u_0=0.8$ in figure),
and read off the corresponding lattice bare couplings $\betaL$,
which are used to define the scale $M$ at each
theory with $N_f$ flavors.}
\label{Fig:u0_const}
\end{center}
\end{figure}

Before going to details, we first explain the basic idea
which follows the FRG analysis by Braun and Gies~\cite{BraunGies}. 
They used the $\tau$ lepton mass $m_{\tau} = 1.777$ (GeV)
as an $N_f$ independent UV reference scale for
theories with any number of flavors.
The initial condition of the renormalisation flow
has been specified via the strong coupling constant
in an $N_f$ independent way:

\begin{align}
\alpha_s(\mu = m_{\tau}) = 0.322\quad
\text{for}\quad {}^{\forall}N_f\ .\label{eq:ini_FRG}
\end{align}

Starting from the common initial condition Eq.~(\ref{eq:ini_FRG}),
 the $N_f$ dependence of the critical temperature $T_c(N_f)$
emerges from 
the 
$N_f$ dependent 
renormalisation flow
at the chiral phase transition scale
$\mu\sim\Lambda_{\mathrm{QCD}}\ll m_{\tau}$.

The $N_f$ dependence of $T_c$
as well as its novel non-analytic behaviour
in the pre-conformal region
becomes free from the choice of
the reference scale~\cite{BraunGies}
by using an $N_f$ independent UV reference scale
much larger than $T_c$.

In order to determine the reference coupling $\gLR$
we utilise our plaquette results $\langle P \rangle$ (equivalently,
the tadpole factor $u_0 = \langle P \rangle^{1/4}$)
shown in Fig.~\ref{Fig:u0_const}. 

Let us consider a constant $u_0$,
for instance $u_0 = 0.8$ in figure,
and read off the corresponding bare lattice couplings at each $N_f$.
The obtained $\gL(N_f)$ 
is used as a reference coupling $\gLR$
and the corresponding mass scale $M(\gLR)$
is again computed according to two loop scaling.

Some remarks on the aforementioned scale setting are in order:
First, we recall the scale setting procedure in the potential scheme,
where the measured normalised force $r^2F(r)$
is proportional to the renormalised coupling $\bar{g}$,
and the specification $\bar{g}^2\propto r_X^2F(r_X) = {}^\exists X$
sets a scale $r_X^{-1}$.
In short, we use our $u_0$ (or equivalently plaquettes) 
to define $\bar{g}$, 
and $u_0 = X$ is regarded as the analog of
the potential scheme scale setting.
Second,
in the leading order of the perturbative expansion,
the renormalised coupling is $N_f$ independent, and
proportional to the Wilson loop~\cite{Wong:2005jx} --
a property that we have already exploited in 
the E-scheme calculation.
Hence the use of an $N_f$ independent $u_0$
approximately gives an $N_f$ independent scale setting,
similarly to
the FRG scale setting method Eq.~(\ref{eq:ini_FRG}).
And third, such an $N_f$ independent scale setting
can be performed in a sufficiently UV regime
$T_c(N_f) \ll M(\gLR)$
by adjusting the value of $u_0$ to satisfy
the condition $\gLR \ll \gTC({}^{\forall}N_f)$.

\begin{figure}
\begin{center}
\includegraphics[width=10cm]{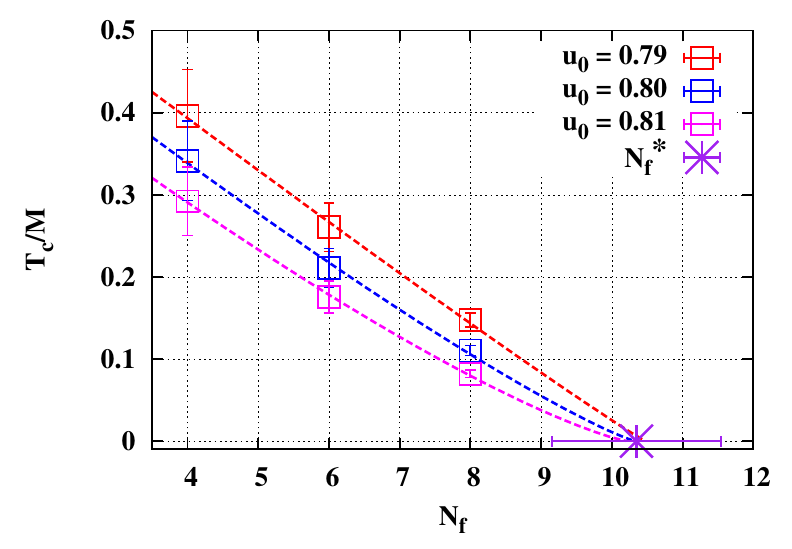}
\caption{
Left:~
The $N_f$ dependence of $T_c/M$ where
$M$ is determined to be a UV scale corresponding to
$u_0=0.79$ (red box),
$0.80$ (blue $\bigcirc$), and
$0.81$ (magenta triangle)
at each theory with $N_f$.
The dashed lines represent fits for data
by assuming the expression Eq.~(\protect\ref{eq:BG_scaling})}
\label{Fig:TcM}
\end{center}
\end{figure}

Note that the 
coupling at the lattice cutoff $a^{-1}(\gLC)$
is $N_t\gg 1$ times larger than $T_c$.
Then, the scale hierarchy $T_c(N_f) \ll a^{-1}(\gLC(N_f))$
allows us to consider a reference scale
much larger than critical temperature
but smaller than the lattice cutoff
$T_c(N_f) \ll M(\gLR) < a^{-1}(\gLC(N_f))$.
We find that $u_0\sim 0.8$ meets this requirement.

In summary, 
the use of $\gLR$ given by $u_0\sim 0.8$
is analogous to
the FRG scale setting method Eq.~(\ref{eq:ini_FRG}),
and is suitable for studying the vanishing
of the critical temperature
by utilising $T_c/M(\gLR)$.

Fig.~\ref{Fig:TcM}
displays the $N_f$ dependence of $T_c/M(\gLR)$
for $u_0 = 0.79$, $0.80$, and $0.81$.
Fitting the data points for $T_c/M(\gLR)$ at $N_f \geq 4$
by using the FRG motivated ansatz,

\begin{align}
T_c= K|N_f^* - N_f|^{(-2b_0^2/b_1)(N_f^*)}
\ ,\label{eq:BG_scaling}
\end{align}

where $b_{0,1}$ has been defined in Eqs.~(\ref{eq:b0}) and (\ref{eq:b1}),
the lower edge of the conformal window is estimated as:
$N_f^* = 10.48 \pm 1.01$ ($u_0 = 0.79$),
$N_f^* = 10.34 \pm 0.88$ ($u_0 = 0.80$),
$N_f^* = 10.23 \pm 0.80$ ($u_0 = 0.81$).
The error-bars involve both fit errors and
statistical errors of data.

We have further investigated the stability against
different choices of $u_0$:
$N_f^*$ is relatively stable
within the range $0.79\leq u_0\leq 0.94$.
The scale cannot be pushed further towards
the UV because of discritization errors. 
On the other hand, 
a small $u_0 \lesssim 0.7$ leads to $M(\gLR)\sim T_c$ or smaller.
In such a case,
the reference scale $M(\gLR)$ 
is affected by infra-red physics and cannot
be used to study the vanishing of $T_c$.
Despite these limitations,
the window of relative stability is however reasonably large,
and suffices to define
an average value for $N_f^*$.
We quote the average among
the three results obtained for $u_0=(0.79,0.80,0.81)$,
i.e. $N_f^* = 10.4 \pm 1.2$.

\section{Learning 
about the Quark Gluon Plasma when studying the threshold
for conformality}\label{sec:QGP}

In this second subsection,
we will follow 
the approach of a recent paper \cite{Liao:2012tw},
and compute the coupling $\gTC (N_f)$ at the scale of
the critical temperature for each $N_f$.
To obtain the coupling $\gTC$ 
at the scale of the temperature,
we evolve the coupling at the scale of the lattice spacing
$a$ up to the temperature inverse scale $N_t a$, still
making use of the two loop scaling, which, as we have seen,
is reasonably well satisfied.

\begin{figure}
\begin{center}
\includegraphics[width=10cm]{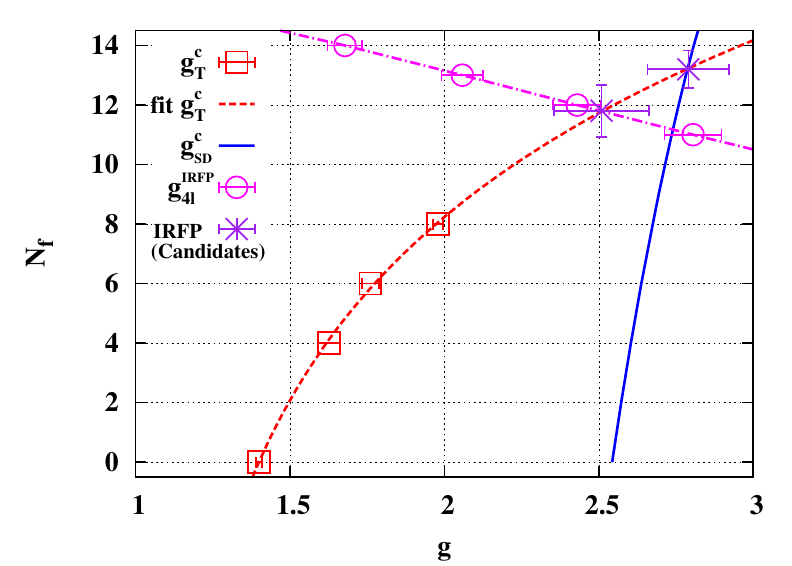}
\caption{
The thermal critical coupling (red $\Box$)
and the fit for them
(dashed red line, with the ansatz Eq.~(\protect\ref{eq:gTC_fit}))
and the values of the zero temperature couplings in the conformal
phase from different estimates, see text for details.
At the critical
number of flavor the thermal critical coupling
should equal the critical
coupling associated with the IRFP.
The procedure is motivated by a recent
study by Shuryak in Ref.~\protect\cite{Liao:2012tw}.}
\label{Fig:gTC}
\end{center}
\end{figure}

The red ($\Box$) symbol in
Fig.~\ref{Fig:gTC} shows $\gTC$ as a function of $N_f$.
We superimpose a fit obtained by using the ansatz proposed in
Ref. \cite{Liao:2012tw}

\begin{align}
{N_f(\gTC) = A\cdot \log~
\bigl[B\cdot(\gTC- \gTC|_{N_f=0}) + 1\bigr]\ .\label{eq:gTC_fit}}
\end{align}

with $A$ and $B$ fit parameters, which describes well the data. 

Since the critical temperature is
zero in the conformal phase,
the thermal critical coupling $\gTC$
should equal a {\em zero temperature} critical coupling $g^c$
when $N_f = N_f^*$.
Of course, $g^c$ is not known exactly
and we have to rely on approximations. 

The first estimate is based on
the best available value $\gSD$
obtained by using
the two-loop Schwinger-Dyson equation \cite{Appelquist}.
In this case,
the lower edge of the conformal window $N_f^*$
is defined by the condition $\gTC(N_f^*) = \gSD(N_f^*)$.
In Fig.~\ref{Fig:gTC} $\gSD$ is plotted as a blue solid line. 
We then estimate the intersection of $\gTC$ and $\gSD$ --
hence the onset of the conformal window
as well as the IRFP coupling at $N_f^*$ -- 
at $(g^*,N_f^*) = (2.79,13.2)\pm (0.13,0.6)$.

One second possibility for estimating $N_f^*$ is the following:
the conformal phase would emerge
when the coupling at IRFP
($g^{\mathrm{IRFP}}$) is not strong enough 
to break chiral symmetry, {\em i.e.}
$g^{\mathrm{IRFP}} \leq \gTC$.
Here, we utilise the four-loop result for $\gIRFP$
\cite{Ryttov:2012nt} as the best available.
In Fig.~\ref{Fig:gTC}, we show 
$\gIRFP$ as magenta $\bigcirc$,
with superimposed a linear interpolation.
In the plot,
we use the results for $\gIRFP$ in the $\bar{\text{MS}}$ scheme.
The errors are estimated by considering the scheme 
dependence \cite{Ryttov:2012nt}, which turns out to be rather mild
at four loops.
We can then locate the intersection of $\gTC$ and $\gIRFP$
and obtain $(g^*,N_f^*) = (2.51,11.8)\pm (0.15,0.9)$.

Ideally,
the three lines in Fig. \ref{Fig:gTC}
should meet at a (single) IRFP fixed point, if all the
quoted results -- including the analytic ones -- were exact.
Indeed 
the intersections we have estimated are consistent 
within the largish errors.
We then quote the average of the above two estimates
as our final result from this analysis, $N_f^*\sim 12.5\pm 1.6$.

In addition, 
we note that $\gTC$ is an increasing function
of $N_f$. This indicates that the quark-gluon plasma
is more strongly coupled at larger $N_f$,
as discussed in Ref.~\cite{Liao:2012tw}.
In turn, this observation might provide a clue into the
nature of the strongly interactive quark gluon plasma.

\section{Two scales?}\label{sec:scale}
Let us elaborate on the circumstance that 
$T_c/\Lambda$ computed using different schemes
($\Lambda = \Lambda_{\mathrm{L}}$ or $\Lambda_{\mathrm{E}}$)
consistently shows an increase with $N_f$, as initially
noted in \cite{Miura:2011mc}.
As discussed in \cite{Miura:2011mc} this indicates that 
$\Lambda_{\mathrm{L/E}}$ vanishes faster than $T_c$
upon approaching the critical number of flavor. Within the various
uncertainties discussed here, this can be taken as a qualitative
indication of scale separation close to the critical
number of flavors.

In the Section above, 
we have estimated the onset of the conformal phase $N^∗_f$ via the vanishing
of $T_c (N_f)/M$. As a next step, it is preferable to define 
$T_c (N_f)/M$ without recourse to perturbation theory. 

\begin{figure}
\begin{center}
\includegraphics[width=10cm]{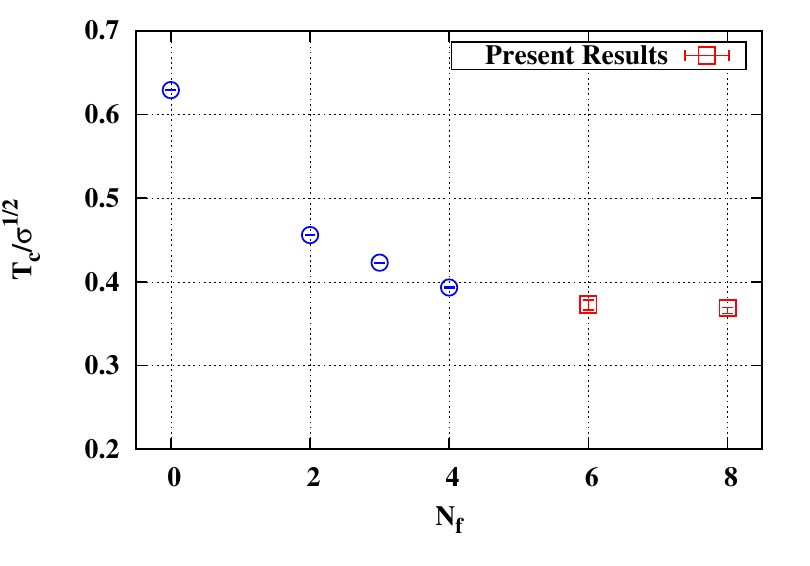}
\caption{
The $T_c/\sqrt{\sigma}$ as a function of $N_f$.
The symbol $\Box$ (red) represents the present results ($N_f = 6,8$).
For a comparison, we have quoted the $T_c/\sqrt{\sigma}$ from
\protect\cite{Laermann:1996xn} ($N_f = 0$),
\protect\cite{Karsch:2000kv} ($N_f = 2,3$),
\protect\cite{Engels:1996ag} ($N_f = 4$),
shown as $\bigcirc$ (blue) symbols.
}\label{Fig:Tc_s}
\end{center}
\end{figure}

To this end, we have adopted the string tension $\sigma$ as a reference scale $M$,
and investigated $T_c / \sqrt{\sigma}$ (Fig.~\ref{Fig:Tc_s}).
The $\sigma$ is evaluated from the Wilson loop
measured on zero temperature lattices, for the
same set of pseudocritical couplings we have identified in our thermal study.
The $T_c/\sqrt{\sigma}$ 
remains stable in the error, again suggesting that our results are a reasonable
approximation of the continuum ones. 

$T_c/\sqrt{\sigma}$ = 0.373(2)(+5,-6) (0.369(4)(+1,-5)) for $N_f = 6(8)$, 
and the decreasing trend becomes less apparent with increasing 
$N_f$ , and $T_c /\sqrt{\sigma}$ does not seem to 
intercept the $N_f$ axis before the asymptotic freedom is lost ($N_f = 16.5$). 
This may not be surprising. We find at least two reasons for the 
non-vanishing $T_c / \sqrt{\sigma}$ : First, $\sigma$ would not
be a ``UV'' quantity and may also be vanishing when
a conformal phase sets in. In other 
words, our result indicates that the regulator of
$T_c$ has to be more UV than $\sigma$ to elucidate the
vanishing of chiral symmetry breaking via $T_c$ . From
this point of view, a quantity $T_c w_0$ where $w_0$ is a
UV scale \cite{Borsanyi:2012zs} defined by the Wilson flow \cite{Luscher:2009eq}
may be a candidate which we are currently evaluating. Suppose
$T_c w_0$ displayed the expected hints of singularity at our
estimated $N_f^* $: our estimate of the critical number of
flavor would be confirmed, and we will have a significant evidence
of a scale separation - the two different scales bing $\sigma$ and $w_0$.
Again however one might argue 
that a finite bare fermion mass breaks the conformality, 
and both $T_c$ and $\sigma$ 
could be defined and finite even in the region $N_f \geq N^∗_f$. 
Thus bare fermion mass
effects to $T_c /\sqrt\sigma$ should be further studied in future.

As indicated in Ref. \cite{Gursoy:2010fj}, 
the ratio $T_c /\sqrt{\sigma}$ is one of the input parameters to set a scale in
models based on the gauge/gravity duality at finite T. 
Such inputs for the (would-be) walking
regime $N_f = 6$ and $N_f = 8$ 
are now available by the present study.

A final caveat concerns the occurrence of a small oscillatory behavior
in the effective mass of the Wilson loop 
with smearing for $N_f=8$. It remains to be seen how these oscillations
relate to the bulk transition observed in the conformal window: 
for instance these observations might confirm the original scenario in which
the bulk transition would still manifest itself in the QCD phase, 
as a (pseudo)singularity unrelated with the chiral transition. 

\section{Summary}\label{sec:summary}

We have presented an overview of some of our results on the phases of QCD
at large number of flavor $N_f$, with some emphases on the scales of the theory, and on
the scale setting procedure: oversimplifying,
QCD generates one dynamically relevant scale for
a small number of flavors, becomes a multi-scale theory when approaching
the conformal window, and then looses its infra red scale.

We need consistent procedures of scale setting in order 
to properly appreciate these phenomena. We are confident that we have taken
at least some steps towards this goal and 
we hope that the strategies we have developed will help further sharpen 
some of the still semi-quantitative estimates presented here.

For $N_f=12$, our measurements of the order parameter
and of the spectrum to our results provide evidence 
towards the existence of a symmetric, 
Coulomb-like phase on the weak coupling side of the lattice bulk transition.  
In the scenario of Refs.~\cite{Appelquist,Miransky:1997},
such a Coulomb-like
region must be entangled to the presence of a conformal IRFP
for the theory with twelve flavors at a continuum limit.
We have then analyzed the spectrum results as a function of mass, and found
them to be well described by power-law fits with a mass anomalous dimension
consistent with other lattice results and
as well as the analytic estimates \cite{Itou:2013ofa}.

On the QCD side ($N_f < N_f^*$), we have investigated 
the chiral phase transition/crossover
with $N_f = 0$ (quenched), $4$, $6$, and $8$.
We have discussed the possible implication for the
(pre-)conformal dynamics at large $N_f$,
and estimated, in a few independent ways,
the number of flavor $N_f^*$:
We have estimated the $N_f^*$ from the vanishing thermal scaling
by extrapolating our critical couplings $\gLC$ to
larger $N_f$ . This gives $N_f^*\sim 11.1\pm 1.6$.
We have extracted
a typical interaction strength $\gTC$
at the scale of critical temperature $T_c$
by utilising our $\gLC$ and the two-loop beta-function,
and compared $\gTC$ to
the zero temperature critical couplings ($\gSD$)
estimated by the two-loop Schwinger-Dyson equation
\cite{Appelquist}
as well as the IRFP position
($\gIRFP$)
of the four-loop beta-function \cite{Ryttov:2012nt}.
The coincidence between $\gTC$ and $\gSD$ or
$\gIRFP$ indicates
the vanishing critical temperature
with the emergence of the conformal phase.
Based on this reasoning,
we have estimated the onset of the conformal window
as $N_f^*\sim 12.5\pm 1.6$.
We have also confirmed
the increasing of $\gTC$ at larger $N_f$
which has been discussed in Ref.~\cite{Liao:2012tw}
and indicates
more strongly interacting non-Abelian plasma at larger $N_f$.

Further, we have examined the $N_f$ dependence of $T_c/M$
for a variety of choices for a reference scale $M$:
we have first considered a UV reference scale $M$
which is determined by utilising the tadpole factor $u_0$.
Then, $T_c/M$ turns out to be a decreasing function of $N_f$
consistently to the FRG observations~\cite{BraunGies},
and the vanishing $T_c/M$ indicates the emergence of the conformal window
around $N_f^* \sim 10.4 \pm 1.2$.

Then we have studied $T_c/\sqrt{\sigma}$ and we are currently extending
our study to $T_c w_0$ : the comparison among these different scale
setting procedures allows a controlled observation of the genuine
singularities - if any - associated with the onset of conformality,
and should highlight the emergence of different scales in the
pre-conformal window. 

Last but not the least, we expect that
our {\em thermodynamic} lattice study for
the large $N_f$ non-Abelian gauge theory
plays an important role as a new connection
between the lattice and the Gauge/Gravity duality
\cite{Gursoy:2010fj,Alho:2012mh}.

\section{Acknowledgments}
MPL and KM were partially supported by the 
PRIN `Frontiers of Strong Interactions' funded by the MIUR. 
This work was in part based on the MILC Collaboration’s 
public lattice gauge theory code~\cite{MILC}. 
The numerical calculations for the chiral phase transition BG/P at CINECA in Italy 
and the Hitachi SR-16000 at YITP, Kyoto University in Japan. 
The numerical calculations for making the gauge
configurations at zero temperature were performed in BG/Q at CINECA in Italy. The Wilson loop
measurements were carried out on the high-performance computing system ϕ at KMI, Nagoya
University in Japan. 
We wish to thank Marc Wagner for providing us the code for the 
Wilson loop measurements. 

\vskip 3 truecm
{\em Note added : After the acceptance of this work we have completed
our paper  {\em On the particle spectrum and the conformal
window}\cite{Lombardo:2014pda},
which addresses the open issues of Section II providing a
nonperturbative
determination of the fermion mass anomalous dimension $\gamma =
0.235(46)$}.

\end{document}